\documentclass[11pt]{article}

\usepackage{hyperref,bm,amsmath,amssymb,a4wide,graphicx,subfigure,wasysym,amsfonts,cite}

\begin{document}

\title{\textbf{Generation of strong magnetic fields in dense quark matter driven
by the electroweak interaction of quarks}}

\author{Maxim Dvornikov$^{a,b,c}$\thanks{maxdvo@izmiran.ru}
\\
$^{a}$\small{\ Pushkov Institute of Terrestrial Magnetism, Ionosphere} \\
\small{and Radiowave Propagation (IZMIRAN),} \\
\small{108840 Troitsk, Moscow, Russia;} \\
$^{b}$\small{\ Physics Faculty, National Research Tomsk State University,} \\
\small{36 Lenin Avenue, 634050 Tomsk, Russia} \\
$^{c}$\small{\ II. Institute for Theoretical Physics, University of Hamburg,} \\ 
\small{149 Luruper Chaussee, D-22761 Hamburg, Germany}}

\date{}

\maketitle

\begin{abstract}
We study the generation of strong large scale magnetic fields in dense
quark matter. The magnetic field growth is owing to the magnetic field
instability driven by the electroweak interaction of quarks. We discuss
the situation when the chiral symmetry
is unbroken in the degenerate quark matter.
In this case we predict the amplification of the seed magnetic
field $10^{12}\,\text{G}$ to the strengths $(10^{14}-10^{15})\,\text{G}$.
In our analysis we use the typical parameters of the quark matter
in the core of a hybrid star or in a quark star. We also discuss the
application of the obtained results to describe the magnetic fields
generation in magnetars.
\end{abstract}



\section{Introduction}

The origin of strong magnetic fields $B\sim10^{15}\,\text{G}$ in
some compact stars, called magnetars~\cite{MerPonMel15}, remains
an open problem of modern astrophysics. Despite the popularity of
some models describing the generation of such magnetic fields, which
are based on magnetohydrodynamics of stellar plasmas, none of them
can satisfactory describe the observational data. These models are
reviewed in Ref.~\cite{MerPonMel15}.

Recently, the methods of elementary particle physics, mainly the chiral
magnetic effect (CME)~\cite{MirSho15}, were applied in Ref.~\cite{ChaZhi10}
to generate toroidal magnetic fields in a neutron star (NS), and in particular
to solve the problem of magnetars~\cite{SigLei16}. The major motivation to apply CME
to produce magnetic fields in NS is that the nonzero chiral
imbalance of electrons $\mu_{5}=\left(\mu_{\mathrm{R}}-\mu_{\mathrm{L}}\right)/2$
is created in nonequilibrium Urca processes, which are parity violating.
It happens since ultrarelativistic left electrons are washed out from
the system producing $\mu_{5}>0$.
The nonzero $\mu_{5}$ generates the electric current of ultrarelativistic electrons along the magnetic field.
This current, in its turn, leads to the magnetic field instability resulting in the growth of
a magnetic field.

Another possibility to utilize the electroweak interaction for the production
of the magnetic field instability was proposed in Refs.~\cite{BoyRucSha12,DvoSem04}.
It consists in the fact that the induced anomalous electric current along the magnetic
field gets the contribution proportional to the difference of the
effective potentials of the effective electroweak interaction of left
and right electrons with background matter. Thus the electroweak interaction
becomes a constant driver of the magnetic field instability. Then
in Refs.~\cite{DvoSem15a,DvoSem15b,DvoSem15c,Dvo16} this idea was
applied to generate strong large scale magnetic fields in NS due to the electroweak electron-nucleon interaction.

The key issue in the application of CME to generate a stellar
magnetic field is the presence of left and right charged fermions
in a star. Strictly speaking, the possibility to separate a fermionic
field into left and right chiral projections is only possible if this
particle is massless, i.e. when the chiral symmetry in unbroken. Despite
the typical energy of an electron in the NS matter is much
greater than its mass, one cannot claim these electrons are chiral particles
there. Thus we can expect that CME for electrons is unlikely to appear in NS.
This claim is also true with respect to the model in Refs.~\cite{DvoSem15a,DvoSem15b,DvoSem15c,Dvo16}.
Note that, for the first time the fact that a nonzero particle mass
destroys CME was noticed in Ref.~\cite{Vil80}. Recently, in Ref.~\cite{Dvo16b},
this result of Ref.~\cite{Vil80} was confirmed in the presence of
the electroweak interaction.

Despite of the above disappointing observation, we can still expect
the existence of astrophysical media where the chiral symmetry is
unbroken. It is the quark matter in the core of a hybrid star (HS) or in
a hypothetical quark star (QS). HS is a NS having the quark core.
QS is based on the strange matter hypothesis. The properties
of these compact stars are reviewed in Ref.~\cite{Gle00}. Note that,
despite of the sporadic claims of the observations of HS/QS (see, e.g., Ref.~\cite{Dai16}), there is a certain skepticism
on the existence of these compact stars.

The present work is devoted to the application of the methods of Refs.~\cite{DvoSem15a,DvoSem15b,DvoSem15c,Dvo16}
to describe the magnetic field instability, leading to its growth,
in quark matter in HS/QS. In Sec.~\ref{sec:BFIELDEVOL},
we derive the kinetic equations describing the evolution of the magnetic
field and chiral imbalances in quark matter. We also formulate the initial conditions
corresponding to a typical astrophysical medium. Then, in Sec.~\ref{sec:RES},
we present the results of the numeric solutions of these kinetic equations.
Finally, in Sec.~\ref{sec:DISC}, we discuss the obtained results
and their applicability for modeling magnetic fields in magnetars.
The computation of the helicity flip rates of quarks in their mutual
collisions is provided in Appendix~\ref{sec:HELFLR}.

\section{Basic equations for the magnetic field evolution in quark matter\label{sec:BFIELDEVOL}}

In this section we shall derive the equations for the evolution of
the spectra of the magnetic helicity density and the magnetic energy
density as well chiral imbalances in degenerate matter containing
$u$ and $d$ quarks interacting by the parity violating electroweak
forces.

Let us consider a dense quark matter consisting of $u$ and $d$ quarks.
The density of this matter is supposed to be high enough for the
chiral symmetry to be restored. In this case we can take that the quarks
are effectively massless. Recently, in Ref.~\cite{Bra16}, it was shown with
help of lattice simulations that
the chiral symmetry has a tendency to restore in a quark matter at high density.
Therefore we can decompose the quark wave
functions into left and right chiral components, which evolve independently,
and attribute different chemical potentials $\mu_{q\mathrm{L,R}}$,
where $q=u,d$, for each chiral component.

Generalizing the results of Refs.~\cite{DvoSem15a,DvoSem15b}, we
get that, in the external magnetic field $\mathbf{B}$, there is the
induced electric current
\begin{equation}\label{eq:Jind}
  \mathbf{J} = \Pi\mathbf{B},
  \quad
  \Pi = \frac{1}{2\pi^{2}}
  \sum_{q=u,d} e_{q}^{2}
  \left(
    \mu_{q5}+V_{q5}
  \right),
\end{equation}
where $e_{u}=2e/3$ and $e_{d}=-e/3$ are the electric charges of
quarks, $e>0$ is the elementary charge, $\mu_{5q}=\left(\mu_{q\mathrm{R}}-\mu_{q\mathrm{L}}\right)/2$
is the chiral imbalance, $V_{5q}=\left(V_{q\mathrm{L}}-V_{q\mathrm{R}}\right)/2$,
and $V_{q\mathrm{L,R}}$ are the effective potentials of the electroweak
interaction of left and right quarks with background fermions. The
potentials $V_{q\mathrm{L,R}}$ were found in Ref.~\cite{Dvo15}
on the basis of the effective Lagrangian for the $ud$ electroweak
interaction,
\begin{equation}\label{eq:Leff}
  \mathcal{L}_{\mathrm{eff}} =
  -\sum_{q=u,d} \bar{q}
  \left(
    \gamma_{0}^{\mathrm{L}}V_{q\mathrm{L}} +
    \gamma_{0}^{\mathrm{R}}V_{q\mathrm{R}}
  \right)q,
\end{equation}
where
\begin{align}\label{eq:VudLR}
  V_{u\mathrm{L}} = & -\frac{G_{\mathrm{F}}}{\sqrt{2}}n_{d}
  \left(
    1-\frac{8}{3}\xi+\frac{16}{9}\xi^{2}-2|V_{ud}|^{2}
  \right),
  \quad
  V_{u\mathrm{R}}=\frac{G_{\mathrm{F}}}{\sqrt{2}}n_{d}
  \left(
    \frac{4}{3}\xi-\frac{16}{9}\xi^{2}
  \right),
  \nonumber
  \\
  V_{d\mathrm{L}} = & -\frac{G_{\mathrm{F}}}{\sqrt{2}}n_{u}
  \left(
    1-\frac{10}{3}\xi+\frac{16}{9}\xi^{2}-2|V_{ud}|^{2}
  \right),
  \quad
  V_{d\mathrm{R}}=\frac{G_{\mathrm{F}}}{\sqrt{2}}n_{u}
  \left(
    \frac{2}{3}\xi-\frac{16}{9}\xi^{2}
  \right).
\end{align}
Here $\gamma_{0}^{\mathrm{L,R}} = \gamma_{0} \left( 1\mp\gamma^{5} \right) /2$,
$\gamma^{5} = \mathrm{i}\gamma^{0}\gamma^{1}\gamma^{2}\gamma^{3}$,
$\gamma^{\mu} = \left( \gamma^{0},\bm{\gamma} \right)$ are the Dirac
matrices, $G_{\mathrm{F}}=1.17\times10^{-5}\,\text{GeV}^{-2}$ is
the Fermi constant, $\xi=\sin^{2}\theta_{\mathrm{W}}=0.23$ is the
Weinberg parameter, $n_{u,d}$ are the number densities of $u$ and
$d$ quarks, and $V_{ud}=0.97$ is the element of the Cabbibo-Kobayashi-Maskawa
matrix. The matter of the star is supposed to be electrically neutral.
Thus we should have $n_{u}=n_{0}/3$ and $n_{d}=2n_{0}/3$, where
$n_{0}=n_{u}+n_{d}$ is the total number density of quarks in the
star.  Using Eq.~\eqref{eq:VudLR} one obtains that
\begin{equation}
  V_{5u} = \frac{G_{\mathrm{F}}}{2\sqrt{2}}\frac{2n_{0}}{3}
  \left(
    2|V_{ud}|^{2}+\frac{4}{3}\xi-1
  \right),
  \quad
  V_{5d} = \frac{G_{\mathrm{F}}}{2\sqrt{2}}\frac{n_{0}}{3}
  \left(
    2|V_{ud}|^{2}+\frac{8}{3}\xi-1
  \right).
\end{equation}
Assuming that $n_{0}=1.8\times10^{38}\thinspace\text{cm}^{-3}$, we
get that $V_{5u}=4.5\thinspace\text{eV}$ and $V_{5d}=2.9\thinspace\text{eV}$.

Note that in Eqs.~(\ref{eq:Jind}) and~(\ref{eq:Leff}) we do not
account for the $uu$ and $dd$ interactions. However as shown in
Ref.~\cite{Dvo14}, basing on the direct calculation of the two loops
contribution to the photon polarization operator, that such contributions
to the induced current in Eq.~(\ref{eq:Jind}) are vanishing.

Using Eq.~(\ref{eq:Jind}) and the results of Refs.~\cite{DvoSem15b},
we can obtain the system of kinetic equations for the spectra of the
density of the magnetic helicity $h(k,t)$ and of the magnetic energy
density $\rho_{\mathrm{B}}(k,t)$, as well as the chiral imbalances
$\mu_{5u}(t)$ and $\mu_{5d}(t)$, in the form,
\begin{align}\label{eq:systgen}
  \frac{\partial h(k,t)}{\partial t}= &
  -\frac{2k^{2}}{\sigma_{\mathrm{cond}}}h(k,t) +
  \frac{8\alpha_{\mathrm{em}}}{\pi\sigma_{\mathrm{cond}}}
  \left\{
    \frac{4}{9}
    \left[
      \mu_{5u}(t)+V_{5u}
    \right] +
    \frac{1}{9}
    \left[
      \mu_{5d}(t)+V_{5d}
    \right]
  \right\}
  \rho_{\mathrm{B}}(k,t),
  \nonumber
  \displaybreak[1]
  \\
  \frac{\partial\rho_{\mathrm{B}}(k,t)}{\partial t}= &
  -\frac{2k^{2}}{\sigma_{\mathrm{cond}}}\rho_{\mathrm{B}}(k,t) +
  \frac{2\alpha_{\mathrm{em}}}{\pi\sigma_{\mathrm{cond}}}
  \left\{
    \frac{4}{9}
    \left[
      \mu_{5u}(t)+V_{5u}
    \right] +
    \frac{1}{9}
    \left[
      \mu_{5d}(t)+V_{5d}
    \right]
  \right\}
  k^{2}h(k,t),
  \nonumber
  \displaybreak[1]
  \\
  \frac{\mathrm{d}\mu_{5u}(t)}{\mathrm{d}t}= & 
  \frac{2\pi\alpha_{\mathrm{em}}}{\mu_{u}^{2}\sigma_{\mathrm{cond}}}
  \frac{4}{9}
  \int\mathrm{d}k
  \bigg[
    k^{2}h(k,t)
    \notag
    \\
    & -
    \frac{4\alpha_{\mathrm{em}}}{\pi}
    \left\{
      \frac{4}{9}
      \left[
        \mu_{5u}(t)+V_{5u}
      \right] +
      \frac{1}{9}
      \left[
        \mu_{5d}(t)+V_{5d}
      \right]
    \right\}
    \rho_{\mathrm{B}}(k,t)
  \bigg] -
  \Gamma_{u}\mu_{5u}(t),
  \nonumber
  \displaybreak[1]
  \\
  \frac{\mathrm{d}\mu_{5d}(t)}{\mathrm{d}t} = & 
  \frac{2\pi\alpha_{\mathrm{em}}}{\mu_{d}^{2}\sigma_{\mathrm{cond}}}
  \frac{1}{9}
  \int\mathrm{d}k
  \bigg[
    k^{2}h(k,t)
    \notag
    \\
    & -
    \frac{4\alpha_{\mathrm{em}}}{\pi}
    \left\{
      \frac{4}{9}
      \left[
        \mu_{5u}(t)+V_{5u}
      \right] +
      \frac{1}{9}
      \left[
        \mu_{5d}(t)+V_{5d}
      \right]
    \right\}
    \rho_{\mathrm{B}}(k,t)
  \bigg] -
  \Gamma_{d}\mu_{5d}(t),
\end{align}
where $\Gamma_{u,d}$ are the rates for the helicity flip in $ud$
plasma, $\alpha_{\mathrm{em}}=e^{2}/4\pi=7.3\times10^{-3}$ is the
QED fine structure constant, $\sigma_{\mathrm{cond}}$ is the electric
conductivity of $ud$ quark matter, and $\mu_{u,d}=\left(3\pi^{2}n_{u,d}\right)^{1/3}$
are the mean chemical potentials of $u$ and $d$ quarks. In the electroneutral
$ud$ plasma, we get that $\mu_{u}=(1/3)^{1/3}\mu_{0}=0.69\mu_{0}$
and $\mu_{d}=(2/3)^{1/3}\mu_{0}=0.87\mu_{0}$, where $\mu_{0}=\left(3\pi^{2}n_{0}\right)^{1/3}=346\,\text{MeV}$.

The functions $h(k,t)$ and $\rho_{\mathrm{B}}(k,t)$ in Eq.~(\ref{eq:systgen})
are related to the total magnetic helicity $H(t)$ and the magnetic
field strength by
\begin{equation}\label{eq:hdef}
  H(t)= \int \mathrm{d}^3 x
  \left(
    \mathbf{A} \cdot \mathbf{B}
  \right) =
  V\int h(k,t)\mathrm{d}k,
  \quad
  B^{2}(t)=2\int\rho_{\mathrm{B}}(k,t)\mathrm{d}k,
\end{equation}
where $V$ is the normalization volume. The integration in Eq.~(\ref{eq:hdef})
is over all the range of the wave number $k$ variation. Note that
we assume the isotropic spectra in Eq.~(\ref{eq:hdef}).

In our model for the magnetic field generation in magnetars, we suggest
that background fermions are degenerate. Nevertheless there is a nonzero
temperature $T$ of the quark matter, which is much less than the
chemical potentials: $T\ll\mu_{q}$. The conductivity of the degenerate
quark matter was estimated in Ref.~\cite{HeiPet93} as
\begin{equation}\label{eq:sigmaQCD}
  \sigma_{\mathrm{cond}} = 4.64\times10^{20}
  \left(
    \alpha_{s}\frac{T}{T_{0}}
  \right)^{-5/3}
  \left(
    \frac{\mu_{0}}{300\,\text{MeV}}
  \right)^{8/3}\text{s}^{-1},
\end{equation}
where $\alpha_{s}$ is the QCD fine structure constant, $T_{0} = (10^{8}-10^{9})\,\text{K}$
is the initial temperature corresponding to the time $t_{0}\sim 10^{2}\,\text{yr}$,
when the star is already in a thermal equilibrium.
Using Eq.~(\ref{eq:sigmaQCD}), we obtain that
\begin{equation}\label{eq:sigmaT}
  \sigma_{\mathrm{cond}} = \sigma_{0}\frac{T_{0}^{5/3}}{T^{5/3}},
  \quad
  \sigma_{0}=3.15\times10^{22}\,\text{s}^{-1},
\end{equation}
where we assume that $\alpha_{s}\sim0.1$. Note that $\sigma_{\mathrm{cond}}$
in quark matter is several orders of magnitude less than the conductivity
of electrons in the nuclear matter in NS~\cite{Kel73}.

The volume density of the internal energy of degenerate background
quarks is $\varepsilon_{\mathrm{T}}=\varepsilon_{0}+\delta\varepsilon_{\mathrm{T}}$~\cite{DvoSem15c},
where $\varepsilon_{0}\sim\mu_{q}^{4}$ is the temperature independent part
and $\delta\varepsilon_{\mathrm{T}}=\left[\mu_{u}^{2}+\mu_{d}^{2}\right]T^{2}/2$
is the temperature correction. In Ref.~\cite{DvoSem15c} we suggested
that the growth of the magnetic field is powered by the transmission
of $\delta\varepsilon_{\mathrm{T}}$ to the magnetic energy density $\rho_{\mathrm{B}}=B^{2}/2$.
The energy conservation law in the magnetized $ud$ plasma reads
$\mathrm{d}\left(\delta\varepsilon_{\mathrm{T}}+\rho_{\mathrm{B}}\right)/\mathrm{d}t=0$~\cite{Dvo16}.
Integrating this expression with the appropriate initial condition
one gets
\begin{equation}\label{eq:TBT0rel}
  \left[
    \mu_{u}^{2}+\mu_{d}^{2}
  \right]
  T^{2} + B^{2} =
  \left[
    \mu_{u}^{2}+\mu_{d}^{2}
  \right]
  T_{0}^{2},
\end{equation}
where we assume that initially the thermal energy is greater than
the magnetic energy, which is the case for a young pulsar. Indeed,
if one starts with a seed field $B_{0}=10^{12}\,\text{G}$, one gets
that $\rho_{\mathrm{B}}(t_{0})=1.9\times10^{-4}\,\text{MeV}^{4}$
and $\delta\varepsilon_{\mathrm{T}}(t_{0})=5.5\,\text{MeV}^{4}$. It means
that
\begin{equation}\label{eq:magcool}
  T^{2}=T_{0}^{2}
  \left(
    1-\frac{B^{2}}{B_{\mathrm{eq}}^{2}}
  \right),
\end{equation}
where the equipartition magnetic field can be found from the following expression~\cite{DvoSem15c}:
\begin{equation}\label{eq:Beq}
  B_{\mathrm{eq}}^{2} =
  \left[
    \mu_{u}^{2}+\mu_{d}^{2}
  \right]
  T_{0}^{2} = 1.23\mu_{0}^{2}T_{0}^{2}.
\end{equation}
Note that Eq.~(\ref{eq:magcool}) describes the magnetic cooling,
i.e. the temperature decreasing because of the magnetic field enhancement.
As we will see later, other channels of the star cooling, such as
the neutrino emission~\cite{Yak01}, are negligible on the time scale
of the magnetic field growth in our model. The dependence of the temperature
on the magnetic field is analogous to the quenching of the parameter
$\Pi$ in Eq.~(\ref{eq:Jind}) introduced in Ref.~\cite{DvoSem15c}
(see also Ref.~\cite{quenching}).

Although we suppose that the chiral symmetry is restored in the star
and quarks are effectively massless, there are induced quark masses
due to the interaction with dense matter. The effective masses of
$u$ and $d$ quarks were computed in~Ref.~\cite{Bra92},
\begin{equation}\label{eq:meff}
  m_{u,d}^{2} = \frac{e_{u,d}^{2}}{8\pi^{2}}\mu_{u,d}^{2}.
\end{equation}
Note that the effective
quark masses in Eq.~(\ref{eq:meff}) should be accounted for only in quarks collisions
(see Appendix~\ref{sec:HELFLR}).
It implies the transitions
between left and right particles in their mutual collisions.
The helicity
flip rates $\Gamma_{u,d}$ for each quark types are computed in Appendix~\ref{sec:HELFLR},
\begin{equation}\label{eq:Gammaud}
  \Gamma_{u} =
  2.98\times10^{-10}\mu_{0} =
  1.59\times10^{14}\thinspace\text{s}^{-1},
  \quad
  \Gamma_{d} =
  5.88\times10^{-12}\mu_{0} =
  3.13\times10^{12}\thinspace\text{s}^{-1},
\end{equation}
where we use Eq.~(\ref{eq:mu5identpart}).

Let us introduce the following dimensionless functions:
\begin{equation}
  \mathcal{H}(\kappa,\tau) =
  \frac{\alpha_{\mathrm{em}}^{2}}{2\mu_{0}^{2}}h(k,t),
  \quad
  \mathcal{R}(\kappa,\tau) =
  \frac{\alpha_{\mathrm{em}}^{2}}{k_{\mathrm{min}}\mu_{0}^{2}}
  \rho_{\mathrm{B}}(k,t),
  \quad
  \mathcal{M}_{u,d}(\tau) =
  \frac{\alpha_{\mathrm{em}}}{\pi k_{\mathrm{min}}}
  \mu_{5(u,d)}(t),
\end{equation}
where we assume $k_{\mathrm{min}}<k<k_{\mathrm{max}}$, $k_{\mathrm{min}}=1/R=2\times10^{-11}\,\text{eV}$,
$R=10\,\text{km}$ is the star radius, $k_{\mathrm{max}}=1/\Lambda_{\mathrm{B}}^{(\mathrm{min})}$,
and $\Lambda_{\mathrm{B}}^{(\mathrm{min})}$ is the minimal scale
of the magnetic field, which is a free parameter. Using the dimensionless
parameters,
\begin{equation}
  \kappa=\frac{k}{k_{\mathrm{min}}},
  \quad
  \tau=\frac{2k_{\mathrm{min}}^{2}}{\sigma_{0}}t,
  \quad
  \mathcal{V}_{u,d} =
  \frac{\alpha_{\mathrm{em}}}{\pi k_{\mathrm{min}}}V_{5(u,d)},
  \quad
  \mathcal{G}_{u,d} =
  \frac{\sigma_{0}\Gamma_{u,d}}{2k_{\mathrm{min}}^{2}},
\end{equation}
as well as Eqs.~(\ref{eq:sigmaT}), (\ref{eq:magcool}), and~(\ref{eq:Gammaud}),
we can rewrite Eq.~(\ref{eq:systgen}) in the form,
\begin{align}\label{eq:systdmls}
  \frac{\partial\mathcal{H}(\kappa,\tau)}{\partial\tau} = &
  \left(
    1-\frac{B^{2}}{B_{\mathrm{eq}}^{2}}
  \right)^{5/6}
  \left[
    -\kappa^{2}\mathcal{H}(\kappa,\tau) + 0.22
    \left(
      4
      \left[
        \mathcal{M}_{u}(\tau)+\mathcal{V}_{u}
      \right] +
      \mathcal{M}_{d}(\tau) + \mathcal{V}_{d}
    \right)
    \mathcal{R}(\kappa,\tau)
  \right],
  \nonumber
  \\
  \frac{\partial\mathcal{R}(\kappa,\tau)}{\partial\tau}= &
  \left(
    1-\frac{B^{2}}{B_{\mathrm{eq}}^{2}}
  \right)^{5/6}
  \left[
    -\kappa^{2}\mathcal{R}(\kappa,\tau) + 0.22
    \left(
      4
      \left[
        \mathcal{M}_{u}(\tau)+\mathcal{V}_{u}
      \right] +
      \mathcal{M}_{d}(\tau)+\mathcal{V}_{d}
    \right)
    \kappa^{2}\mathcal{H}(\kappa,\tau)
  \right],
  \nonumber
  \\
  \frac{\mathrm{d}\mathcal{M}_{u}(\tau)}{\mathrm{d}\tau}= &
  1.85
  \left(
    1-\frac{B^{2}}{B_{\mathrm{eq}}^{2}}
  \right)^{5/6}
  \int_{1}^{\kappa_{\mathrm{max}}}\mathrm{d}\kappa
  \big[
    \kappa^{2}\mathcal{H}(\kappa,\tau)
    \notag
    \\
    & -
    0.22
    \left(
      4
      \left[
        \mathcal{M}_{u}(\tau)+\mathcal{V}_{u}
      \right] +
      \mathcal{M}_{d}(\tau)+\mathcal{V}_{d}
    \right)\mathcal{R}(\kappa,\tau)
  \big] -
  \mathcal{G}_{u}\mathcal{M}_{u}(\tau),
  \nonumber
  \\
  \frac{\mathrm{d}\mathcal{M}_{d}(\tau)}{\mathrm{d}\tau}= &
  0.29
  \left(
    1-\frac{B^{2}}{B_{\mathrm{eq}}^{2}}
  \right)^{5/6}
  \int_{1}^{\kappa_{\mathrm{max}}}\mathrm{d}\kappa
  \big[
    \kappa^{2}\mathcal{H}(\kappa,\tau)
    \notag
    \\
    & -
    0.22
    \left(
      4
      \left[
        \mathcal{M}_{u}(\tau)+\mathcal{V}_{u}
      \right] +
      \mathcal{M}_{d}(\tau)+\mathcal{V}_{d}
    \right)
    \mathcal{R}(\kappa,\tau)
  \big] -
  \mathcal{G}_{d}\mathcal{M}_{d}(\tau),
\end{align}
where $\kappa_{\mathrm{max}}=k_{\mathrm{max}}/k_{\mathrm{min}}$,
$B^{2}$ and $B_{\mathrm{eq}}^{2}$ are given in Eqs.~(\ref{eq:hdef})
and~(\ref{eq:Beq}).

While solving of Eq.~(\ref{eq:systdmls}) numerically, we use the initial
Kolmogorov spectrum of the magnetic energy density, $\rho_{\mathrm{B}}(k,t_{0})=\mathcal{C}k^{-5/3}$,
where the constant $\mathcal{C}$ can be obtained by equating the
initial magnetic energy density, computed on the basis of Eq.~(\ref{eq:hdef}),
to $B_{0}^{2}/2$ (see Ref.~\cite{DvoSem15b}). The initial spectrum
of the magnetic helicity density is $h(k,t_{0})=2r\rho_{\mathrm{B}}(k,t_{0})/k$,
where the parameter $0\leq r\leq1$, corresponds to initially nonhelical,
$r=0$, and maximally helical, $r=1$, fields.

In Ref.~\cite{DvoSem15a} we found that the evolution of the magnetic
field is almost independent on the initial values of the chiral imbalances
$\mu_{5(u,d)}(t_{0})$ because of the huge helicity flip rates $\Gamma_{u,d}$.
Therefore we can take almost arbitrary values of $\mu_{5(u,d)}(t_{0})$
only requiring that $\mu_{5(u,d)}(t_{0})\ll\mu_{u,d}$. In our simulations
we shall take that $\mu_{5u}(t_{0})=\mu_{5d}(t_{0})=1\,\text{MeV}$.

\section{Results of the numeric solution of the kinetic equations\label{sec:RES}}

In this section we present the results of the numerical solution of
Eq.~(\ref{eq:systdmls}) with the initial conditions corresponding
to a quark matter in a compact star.

\begin{figure}
  \centering
  \subfigure[]
  {\label{1a}
  \includegraphics[scale=.23]{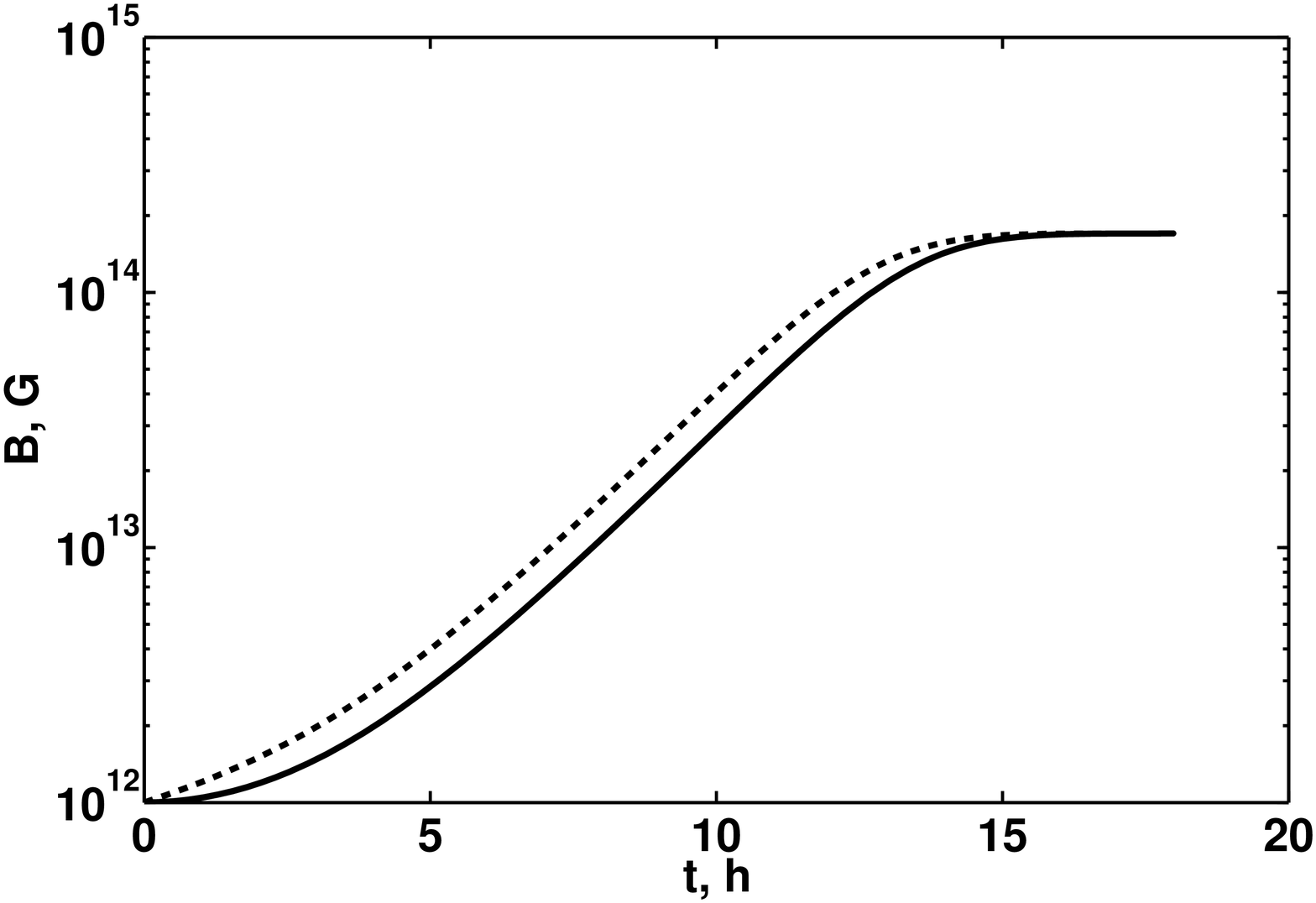}}
  \hskip-.7cm
  \subfigure[]
  {\label{1b}
  \includegraphics[scale=.23]{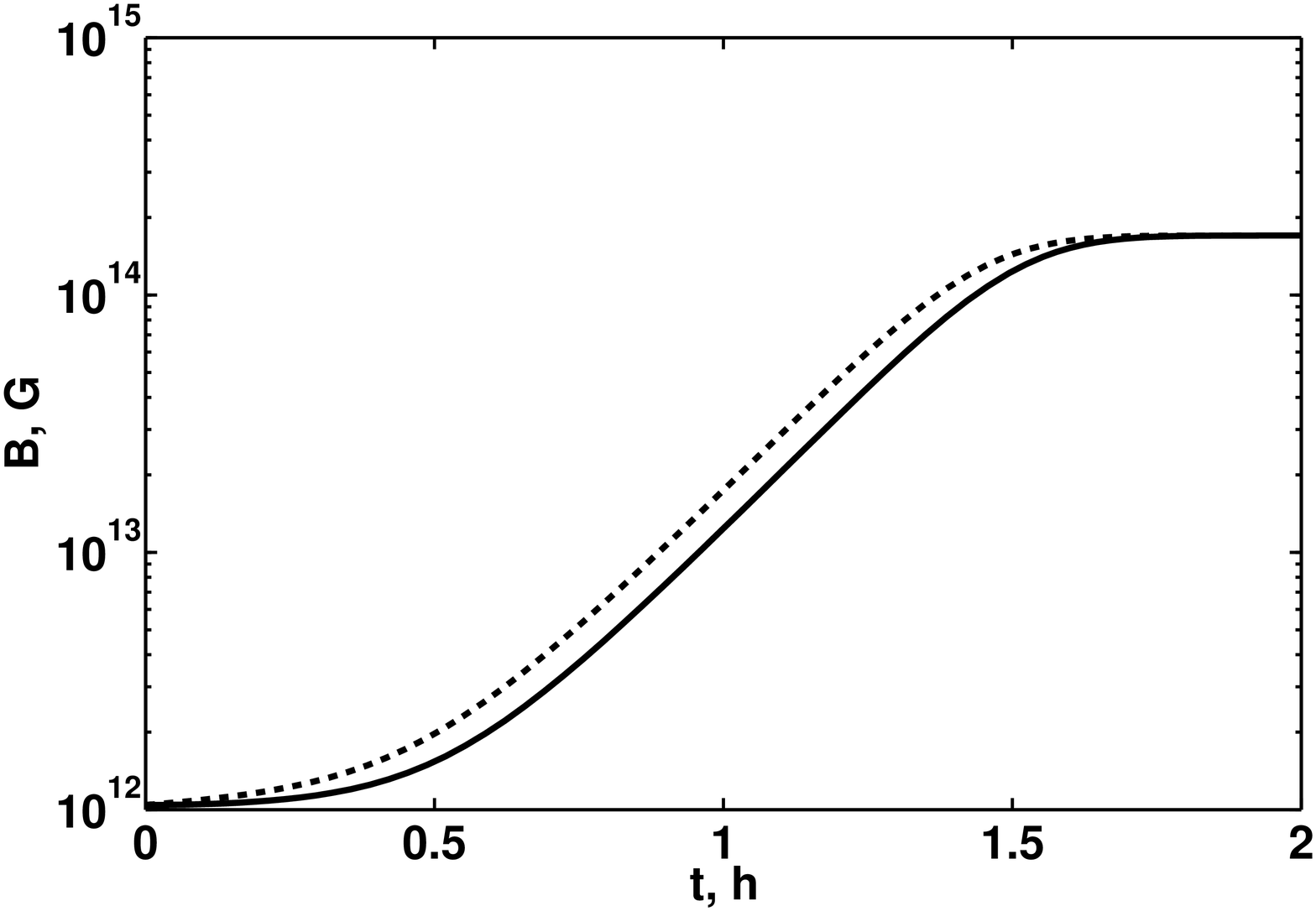}}
  \\
  \subfigure[]
  {\label{1c}
  \includegraphics[scale=.23]{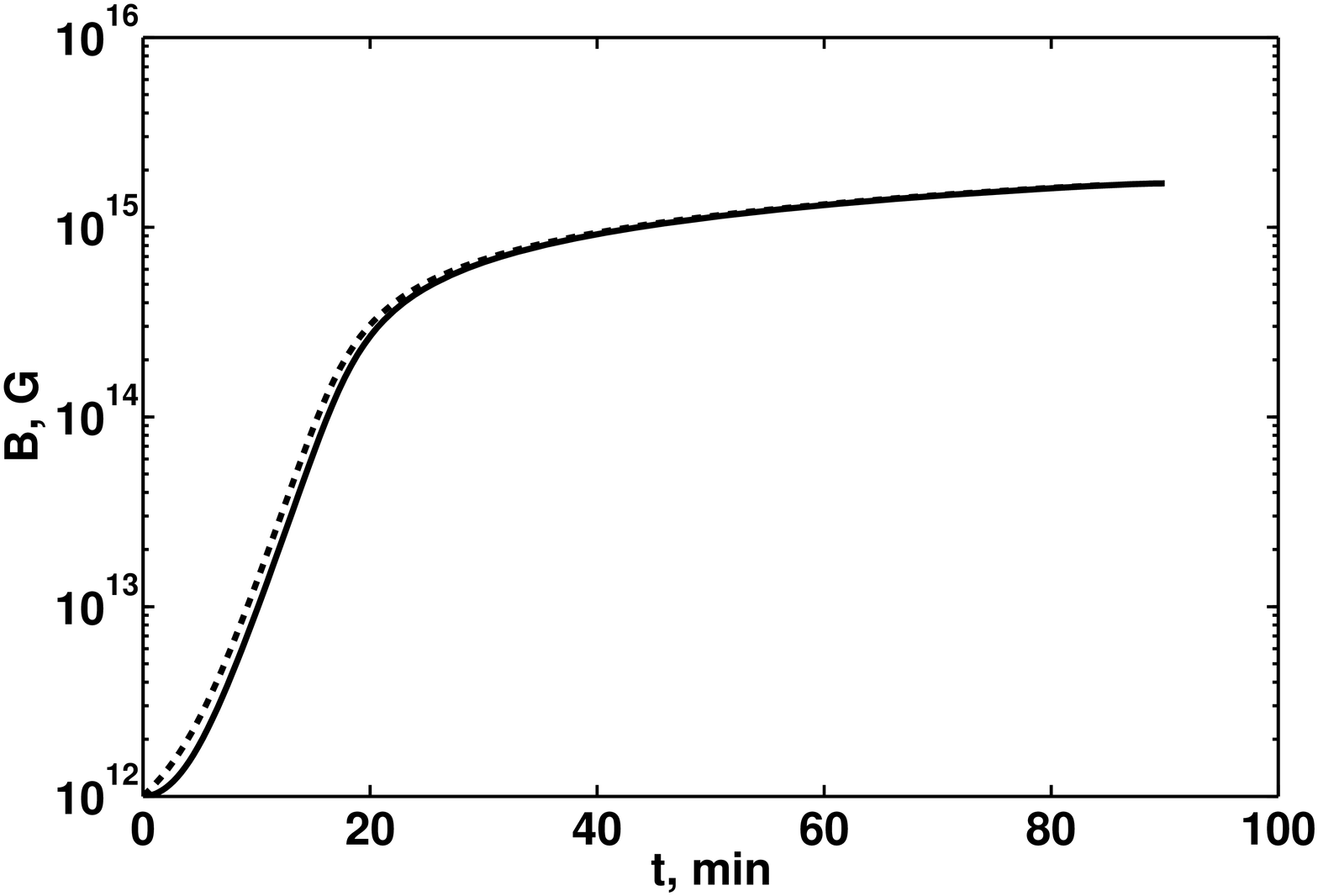}}
  \hskip-.7cm
  \subfigure[]
  {\label{1d}
  \includegraphics[scale=.23]{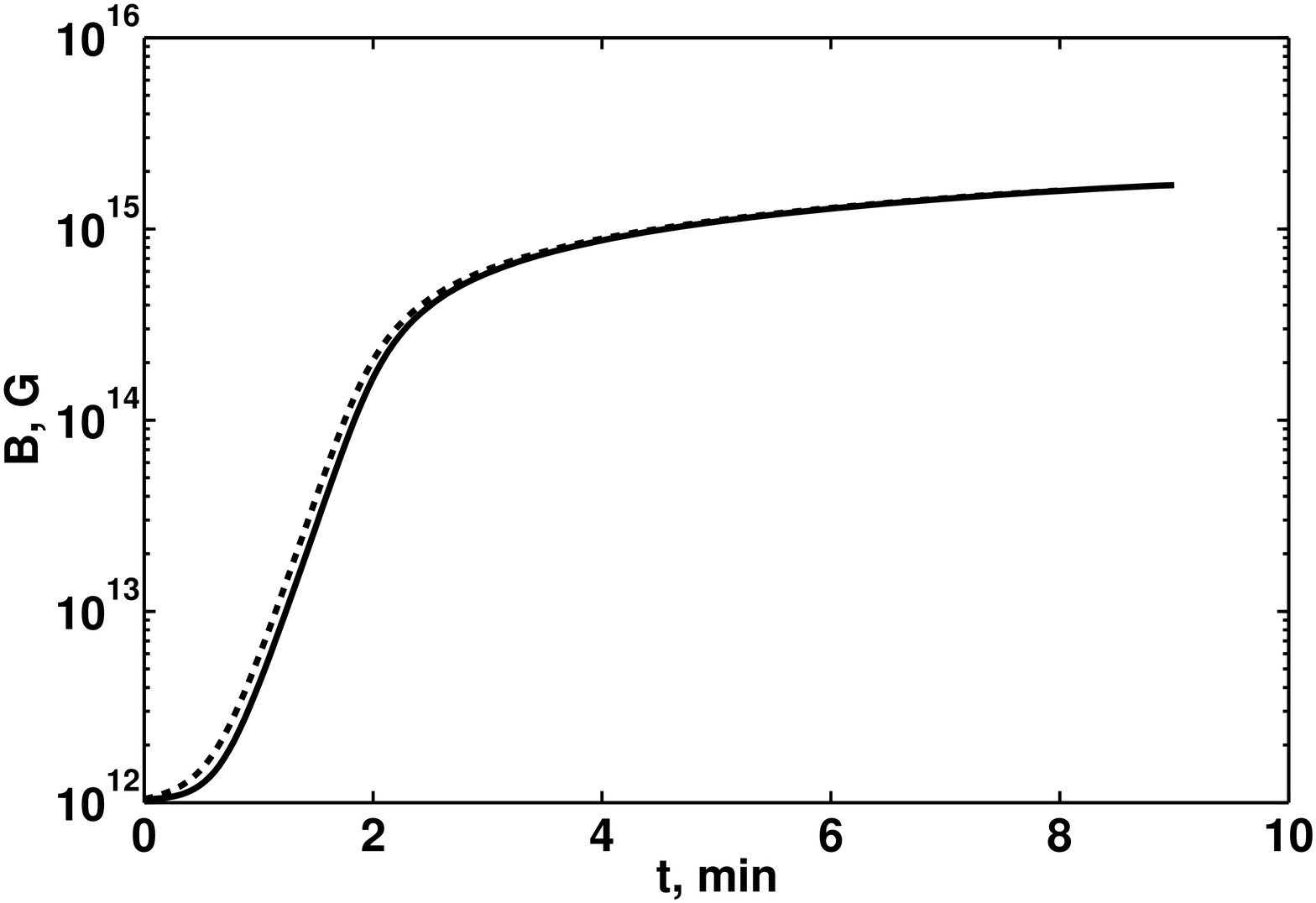}}
  \protect
  \caption{\label{fig:Bevol}
    The magnetic field versus time for different initial temperatures
    $T_{0}$ and minimal length scales   
    $\Lambda_{\mathrm{B}}^{(\mathrm{min})}$.
    The solid lines correspond to initially nonhelical fields with $r=0$
    and dashed ones to the fields having maximal initial helicity, $r=1$.
    (a)~$T_{0}=10^{8}\,\text{K}$ and   
    $\Lambda_{\mathrm{B}}^{(\mathrm{min})}=1\,\text{km}$.
    (b)~$T_{0}=10^{8}\,\text{K}$ and   
    $\Lambda_{\mathrm{B}}^{(\mathrm{min})}=100\,\text{m}$.
    (c)~$T_{0}=10^{9}\,\text{K}$ and 
    $\Lambda_{\mathrm{B}}^{(\mathrm{min})}=1\,\text{km}$.
    (d)~$T_{0}=10^{9}\,\text{K}$ and 
    $\Lambda_{\mathrm{B}}^{(\mathrm{min})}=100\,\text{m}$. 
  }
\end{figure}

In Fig.~\ref{fig:Bevol} we show the amplification of the initial magnetic
field $B_{0}=10^{12}\,\text{G}$ by two or three orders of magnitude.
This result is obtained by numerically solving Eq.~(\ref{eq:systdmls})
with the initial conditions discussed in Sec.~\ref{sec:BFIELDEVOL}.
These initial conditions are quite possible in a dense quark matter
in a HS/QS.

One can see in Fig.~\ref{fig:Bevol} that the magnetic field reaches the saturated strength $B_{\mathrm{sat}}$.
This result is analogous to the findings of Refs.~\cite{DvoSem15c,Dvo16}. For $T_{0}=10^{8}\,\text{K}$
in Figs.~\ref{1a}
and~\ref{1b}, $B_{\mathrm{sat}}\approx1.1\times10^{14}\,\text{G}$;
and for $T_{0}=10^{9}\,\text{K}$ in Figs.~\ref{1c} and~\ref{1d},
$B_{\mathrm{sat}}\approx1.1\times10^{15}\,\text{G}$. However, unlike Refs.~\cite{DvoSem15c,Dvo16},
$B_{\mathrm{sat}}$ in Fig.~\ref{fig:Bevol} is defined entirely
by $T_{0}$. The obtained $B_{\mathrm{sat}}$ is close to the magnetic
field strength predicted in magnetars~\cite{MerPonMel15}, especially
for $T_{0}=10^{9}\,\text{K}$.

The time of the magnetic field growth to $B_{\mathrm{sat}}$ is several
orders of magnitude shorter than in Refs.~\cite{DvoSem15c,Dvo16}. This fact
is due to the smaller value of the electric conductivity $\sigma_{\mathrm{cond}}$
in quark matter in Eq.~(\ref{eq:sigmaQCD}) compared to
$\sigma_{\mathrm{cond}}$ for electrons in nuclear matter which we used in Refs.~\cite{DvoSem15c,Dvo16}.
This fact can be understood with help of the Faraday equation,
\begin{equation}\label{FE}
  \frac{\partial\mathbf{B}}{\partial t} =
  \frac{\Pi}{\sigma_{\mathrm{cond}}}
  \left(
    \nabla\times\mathbf{B}
  \right) +
  \frac{1}{\sigma_{\mathrm{cond}}}\nabla^{2}\mathbf{B},
\end{equation}
which is equivalent to the first two lines in Eq.~(\ref{eq:systgen}). Using Eq.~\eqref{FE}
one gets that the saturation time $t_{\mathrm{sat}}\sim\sigma_{\mathrm{cond}}/\Pi\Lambda_{\mathrm{B}}$,
where $\Lambda_{\mathrm{B}}$ is the magnetic field scale. It means
that the smaller $\sigma_{\mathrm{cond}}$ is, the faster the magnetic
field reaches $B_{\mathrm{sat}}$. Moreover, we can see that short
scale magnetic field should reach $B_{\mathrm{sat}}$ faster. The later
fact, which was also established in Refs.~\cite{DvoSem15b,DvoSem15c,Dvo16},
is confirmed by the comparison of Figs.~\ref{1a}
and~\ref{1b} as well as Figs.~\ref{1c} and~\ref{1d}.

In our model of the magnetic field generation, the thermal energy
of background fermions is converted to the magnetic energy. One can say that a
star cools down magnetically. The typical values of $t_{\mathrm{sat}}$
are $\apprle10\,\text{h}$ in Figs.~\ref{1a} and~\ref{1b}
and $\lesssim10^{2}\,\text{min}$ in Figs.~\ref{1c} and~\ref{1d}.
At such short time scales, other cooling channels, such as that due to the neutrino
emission~\cite{Yak01}, do not contribute to the temperature evolution
significantly. Therefore, unlike Refs.~\cite{DvoSem15b,DvoSem15c,Dvo16},
we omit them in our present simulations.

In Fig.~\ref{fig:Bevol} we can see that, although the initial magnetic
helicity can be different (see solid and dashed lines there), the
subsequent evolution of such magnetic fields is almost indistinguishable,
especially at $t \sim t_{\mathrm{sat}}$. It means that, besides the
generation of strong magnetic field, we also generate the magnetic
helicity in quark matter. This result is in the agreement with Refs.~\cite{DvoSem15b,DvoSem15c,Dvo16}.

\section{Discussion\label{sec:DISC}}

In the present work we have applied the mechanism for the magnetic
field generation, proposed in Refs.~\cite{DvoSem15a,DvoSem15b,DvoSem15c},
to create strong large scale magnetic fields in dense quark matter.
This mechanism is based on the magnetic field instability driven a parity violating electroweak
interaction between particles in
the system. We have established the system of kinetic equations for
the spectra of the magnetic helicity density and the magnetic energy
density, as well as for the chiral imbalances, and have solved it numerically.

Although there is a one-to-one correspondence between the mechanisms
for the magnetic field generation in Refs.~\cite{DvoSem15a,DvoSem15b,DvoSem15c,Dvo16}
and in the present work, the scenario described here is likely to
be more realistic. As mentioned in Ref.~\cite{Dvo16b} the generation
of the anomalous current in Eq.~(\ref{eq:Jind}) is impossible for
massive particles. Electrons in NS are ultrarelativistic
but have a nonzero mass. As found in Ref.~\cite{Rub86}, the chiral
symmetry can be restored at densities $n\sim M_{\mathrm{W}}^{3}\sim10^{46}\,\text{cm}^{-3}$,
that is much higher than one can expect in NS. Therefore
the chiral magnetic effect for electrons as well as the results of
Refs.~\cite{DvoSem15a,DvoSem15b,DvoSem15c,Dvo16} are unlikely to
be applied in NS. Recently this fact was also mentioned in Ref.~\cite{Dvo16b}.

On the contrary, the chiral symmetry was found in Ref.~\cite{DexSch10}
to be restored for lightest $u$ and $d$ quarks even at densities
corresponding to a core of HS or in QS. Accounting
for the existence of the electroweak parity violating interaction
between $u$ and $d$ quarks, we can conclude that the application
of the methods of Refs.~\cite{DvoSem15a,DvoSem15b,DvoSem15c,Dvo16}
to the quark matter in a compact star is quite plausible.

We have obtained that, in quark matter, the seed magnetic field
$B_{0}=10^{12}\,\text{G}$, which is typical in
a young pulsar, is amplified up to $B_{\mathrm{sat}}\sim\left(10^{14}-10^{15}\right)\,\text{G}$,
depending on the initial temperature.
Such magnetic fields are predicted
in magnetars~\cite{MerPonMel15}. Therefore HS/QS can
become a magnetar. The obtained growth time of the magnetic field
to $B_{\mathrm{sat}}$ is much less than that in electron-nucleon
case studied in Refs.~\cite{DvoSem15a,DvoSem15b,DvoSem15c,Dvo16}.
It means that, in our model, strong magnetic fields are generated
quite rapidly with $t_{\mathrm{sat}}\sim$~several hours after a
star is in a thermal equilibrium.

Note that, in the present work, instead of the quenching of the parameter
$\Pi$ in Eq.~(\ref{eq:Jind}) suggested in Ref.~\cite{DvoSem15c}
to avoid the excessive growth of the magnetic field, we used the conservation of the total
energy in Eq.~(\ref{eq:magcool}) and the dependence
of the electric conductivity on the temperature in Eq.~(\ref{eq:sigmaT}); cf. Ref.~\cite{Dvo16}.
It results in a more explicit saturation of the magnetic field in
Eq.~(\ref{eq:systdmls}); cf. Fig.~\ref{fig:Bevol}.

Despite the plausibility of the results, several important assumptions
were made. Firstly, while calculating the helicity flip rates in Appendix~\ref{sec:HELFLR},
we have taken that quarks exchange by plasmons in their scattering.
It is, however, known (see, e.g., Ref.~\cite{KapToi88}) that modified
effective interaction potentials can exist in a dense degenerate matter.
If one takes into account these interactions it can somehow change
the values of $\Gamma_{u,d}$. Nevertheless, since the present work
is a qualitative study of the magnetic field generation in the degenerate
quark matter, we shall restrict ourselves to the the plasmon interaction of
quarks. 

Secondly, we have considered the simplest case of a compact star consisting
of only $u$ and $d$ quarks. However, strange stars, having a certain
fraction of $s$ quarks are also actively studied~\cite[pp.~414--440]{Gle00}.
The nonzero fraction of $s$ quarks, which cannot exceed $1/3$, is
also required by the beta equilibrium. Nonetheless $s$ quarks are
unlikely to contribute significantly to the generation of magnetic
fields in our model. Firstly, the mass of an $s$ quark $m_{s}=150\,\text{MeV}$
is quite great, i.e. the chiral symmetry will remain broken for these
particles. Thus, $s$ quarks do not contribute to the induced current
in Eq.~(\ref{eq:Jind}). Secondly, even if $s$ quarks contribute
to the helicity flip rates of $u$ and $d$ quarks, it will not change
the evolution of the magnetic field. Indeed, $\Gamma_{u,d}$ computed
in Appendix~\ref{sec:HELFLR}, is already great enough to wash out
the initial chiral imbalances $\mu_{5(u,d)}(0)$. Any bigger contribution
to $\Gamma_{u,d}$ will eliminate $\mu_{5(u,d)}(0)$ faster. However,
the growth of the magnetic field is driven by $V_{5(u,d)}$, which
is constant, rather than by $\mu_{5(u,d)}$.

Summarizing, we have described the generation of strong large scale
magnetic fields in dense quark matter driven by the magnetic field
instability caused by the electroweak interaction of quarks. The described
phenomenon may well exist in the core of HS or in QS.
We suggest that the obtained results can have implication to
the problem of magnetars since the generated magnetic fields have
strength close to that predicted in these highly magnetized compact stars.

\section*{Acknowledgements}

I am thankful to S.I.~Blinnikov, A.V.~Borisov, V.V.~Braguta, M.I.~Krivoruchenko, N.~Leite, A.E.~Lobanov,
B.V.~Martemyanov, K.A.~Postnov, V.B.~Semikoz, G.~Sigl, M.I.~Vysotsky, V.I.~Zakharov, and V.Ch.~Zhukovsky
for useful discussions, as well as to the Tomsk State University Competitiveness Improvement Program, RFBR
(research project No.~15-02-00293), and DAAD (grant No.~91610946) for partial support.

\appendix

\section{Helicity flip rates in degenerate quark matter}\label{sec:HELFLR}

In this Appendix we shall compute the helicity flip rates of $u$
and $d$ quarks in their collisions in dense matter as well as derive the kinetic equations for the chiral imbalances.

As mentioned in Sec.~\ref{sec:BFIELDEVOL}, quarks acquire effective
masses in dense matter. Thus the helicity of quarks will change when
the particles collide. There are three types of reactions: (a)~scattering
of identical quarks, with helicities of both particles being changed;
(b)~scattering of different quark flavors, with helicities of both
particles being changed; and (c)~scattering of different quark flavors,
with helicity of only one particle being changed. We shall successively
discuss all the cases. Quarks are supposed to interact by the plasmon
exchange.

\paragraph{Scattering of identical quarks}

There are four reactions in this group: $u_{\mathrm{L}}u_{\mathrm{L}}\to u_{\mathrm{R}}u_{\mathrm{R}}$,
$d_{\mathrm{L}}d_{\mathrm{L}}\to d_{\mathrm{R}}d_{\mathrm{R}}$, $u_{\mathrm{R}}u_{\mathrm{R}}\to u_{\mathrm{L}}u_{\mathrm{L}}$,
and $d_{\mathrm{R}}d_{\mathrm{R}}\to d_{\mathrm{L}}d_{\mathrm{L}}$.
We study in details only the process $u_{\mathrm{L}}(p_{1})+u_{\mathrm{L}}(p_{2})\to u_{\mathrm{R}}(p'_{1})+u_{\mathrm{R}}(p'_{2})$,
where $p_{1,2}^{\mu}=\left(E_{1,2},\mathbf{p}_{1,2}\right)$ are the
momenta of incoming quarks and $p_{1,2}^{\prime\mu}=\left(E'_{1,2},\mathbf{p}'_{1,2}\right)$
are the momenta of outgoing quarks. In this reaction, the number of
left particles is decreased by two units and the number of right particles
is increased by two units. Other reactions in this group can be studied
analogously.

The matrix element has the form,
\begin{equation}\label{eq:MuLuLuRuR}
  \mathcal{M}=\mathrm{i}e_{u}^{2}
  \left[
    \frac{1}{t}
    \bar{u}(p'_{1})\gamma^{\mu}u(p_{1}) \cdot
    \bar{u}(p'_{2})\gamma_{\mu}u(p_{2}) -
    \frac{1}{u}
    \bar{u}(p'_{2})\gamma^{\mu}u(p_{1}) \cdot
    \bar{u}(p'_{1})\gamma_{\mu}u(p_{2})
  \right].
\end{equation}
where $t=\left(p'_{1}-p_{1}\right)^{2}$ and $u=\left(p'_{2}-p_{1}\right)^{2}$
are the Mandelstam variables. The square of the matrix element in
Eq.~(\ref{eq:MuLuLuRuR}) is
\begin{align}\label{eq:M2uLuLuRuR}
  |\mathcal{M}|^{2}= & e_{u}^{4}
  \bigg[
    \frac{1}{t^{2}}
    \text{tr}
    \left(
      \rho'_{2}\gamma_{\mu}\rho_{2}\gamma_{\nu}
    \right)
    \cdot
    \text{tr}
    \left(
      \rho'_{1}\gamma^{\mu}\rho_{1}\gamma^{\nu}
    \right) +
    \frac{1}{u^{2}}
    \text{tr}
    \left(
      \rho'_{1}\gamma_{\mu}\rho_{2}\gamma_{\nu}
    \right)
    \cdot
    \text{tr}
    \left(
      \rho'_{2}\gamma^{\mu}\rho_{1}\gamma^{\nu}
    \right)
    \nonumber
    \\
    &
    -\frac{1}{tu}
    \text{tr}
    \left(
      \rho'_{2}\gamma_{\mu}\rho_{2}\gamma_{\nu}
      \rho'_{1}\gamma^{\mu}\rho_{1}\gamma^{\nu}
    \right) -
    \frac{1}{tu}
    \text{tr}
    \left(
      \rho'_{1}\gamma_{\mu}\rho_{2}\gamma_{\nu}
      \rho'_{2}\gamma^{\mu}\rho_{1}\gamma^{\nu}
    \right)
  \bigg],
\end{align}
where the density matrices are~\cite[pp.~106--111]{BerLifPit82}
\begin{align}\label{eq:densmatr}
  \rho_{1,2} = & \frac{1}{2}
  \left[
    \left(
      \gamma\cdot p_{1,2}
    \right) +
    m_{u}
  \right]
  \left[
    1+\gamma^{5}
    \left(
      \gamma\cdot a_{1,2}
    \right)
  \right],
  \notag
  \\
  \rho'_{1,2} = & \frac{1}{2}
  \left[
    \left(
      \gamma\cdot p'_{1,2}
    \right) +
    m_{u}
  \right]
  \left[
    1+\gamma^{5}
    \left(
      \gamma\cdot a'_{1,2}
    \right)
  \right].
\end{align}
Here $m_{u}$ is the effective mass given in Eq.~(\ref{eq:meff})
and the polarization vectors are~\cite{Dvo16}
\begin{equation}\label{eq:4dpolar}
  a_{1,2}^{\mu} =
  \frac{1}{m_{u}}
  \left(
    -p_{1,2},-E_{1,2}\mathbf{n}_{1,2}
  \right),
  \quad
  a_{1,2}^{\prime\mu}=\frac{1}{m_{u}}
  \left(
    p'_{1,2},E'_{1,2}\mathbf{n}'_{1,2}
  \right),
\end{equation}
which correspond to left and right particles. Here $\mathbf{n}_{1,2}$ and $\mathbf{n}'_{1,2}$ are the unit vectors along $\mathbf{p}_{1,2}$ and $\mathbf{p}'_{1,2}$.

Choosing the center-of-mass frame of colliding quarks and assuming
the elastic scattering, one gets that
\begin{align}\label{eq:truLuLuRuR}
  & \text{tr}
  \left(
    \rho'_{2}\gamma_{\mu}\rho_{2}\gamma_{\nu}
  \right)
  \cdot
  \text{tr}
  \left(
    \rho'_{1}\gamma^{\mu}\rho_{1}\gamma^{\nu}
  \right) =
  16m_{u}^{4}\sin^{4}\frac{\theta_{\mathrm{cm}}}{2},
  \notag
  \\  
  & \text{tr}
  \left(
    \rho'_{1}\gamma_{\mu}\rho_{2}\gamma_{\nu}
  \right)
  \cdot
  \text{tr}
  \left(
    \rho'_{2}\gamma^{\mu}\rho_{1}\gamma^{\nu}
  \right) =
  16m_{u}^{4}\cos^{4}\frac{\theta_{\mathrm{cm}}}{2},
  \nonumber
  \\
  & \text{tr}
  \left(
    \rho'_{2}\gamma_{\mu}\rho_{2}\gamma_{\nu}
    \rho'_{1}\gamma^{\mu}\rho_{1}\gamma^{\nu}
  \right) =
  \text{tr}
  \left(
    \rho'_{1}\gamma_{\mu}\rho_{2}\gamma_{\nu}
    \rho'_{2}\gamma^{\mu}\rho_{1}\gamma^{\nu}
  \right) =
  -4m_{u}^{4}\sin^{2}\theta_{\mathrm{cm}},
\end{align}
where $\theta_{\mathrm{cm}}$ is the scattering angle, i.e. the angle between $\mathbf{p}_{1}$
and $\mathbf{p}'_{1}$ in the center-of-mass frame. In the same
frame one has
\begin{equation}\label{eq:tucm}
  t=-2E_{\mathrm{cm}}^{2}
  \left(
    1-\cos\theta_{\mathrm{cm}}
  \right),
  \quad
  u=-2E_{\mathrm{cm}}^{2}
  \left(
    1+\cos\theta_{\mathrm{cm}}
  \right),
\end{equation}
where $E_{\mathrm{cm}}$ is the energy of colliding quarks in the
center-of-mass frame. In Eq.~(\ref{eq:tucm}), we also assume that the
scattering is elastic. We can express $E_{\mathrm{cm}}$ in term of
the variables in the laboratory frame, i.e. where the star is at rest,
as $E_{\mathrm{cm}}^{2} \approx \left\{ m_{u}^{2}+E_{1}E_{2} \left[ 1- \left( \mathbf{n}_{1}\cdot\mathbf{n}_{2} \right) \right] \right\} /2$.
Since we study the probability in the lowest order in the effective
mass and traces in Eq.~(\ref{eq:truLuLuRuR}) are proportional to
$m_{u}^{4}$, we neglect $m_{u}$ in Eq.~(\ref{eq:tucm}) as well as in
the following calculations.

Finally, Eq.~(\ref{eq:M2uLuLuRuR}) takes the form
\begin{align}\label{eq:M2ushort}
  |\mathcal{M}|^{2}= &
  \frac{16m_{u}^{4}e_{u}^{4}}{
  \left\{ 
    m_{u}^{2}+E_{1}E_{2}
    \left[
      1 -
      \left(
        \mathbf{n}_{1}\cdot\mathbf{n}_{2}
      \right)
    \right]
  \right\} ^{2}}.
\end{align}
The total probability of the process has the form~\cite[pp.~247--252]{BerLifPit82},
\begin{align}\label{eq:WuLuLuRuR}
  W= & \frac{V}{64(2\pi)^{8}}
  \int
  \frac{\mathrm{d}^{3}p_{1}\mathrm{d}^{3}p_{2}
  \mathrm{d}^{3}p'_{1}\mathrm{d}^{3}p'_{2}}
  {E_{1}E_{2}E'_{1}E'_{2}}
  \delta^{4}
  \left(
    p_{1}+p_{2}-p'_{1}-p'_{2}
  \right)
  |\mathcal{M}|^{2}
  \nonumber
  \\
  & \times  
  f
  \left(
    E_{1}-\mu_{u\mathrm{L}}
  \right)
  f
  \left(
    E_{2}-\mu_{u\mathrm{L}}
  \right)
  \left[
    1-f
    \left(
      E'_{1}-\mu_{u\mathrm{R}}
    \right)
  \right]
  \left[
    1-f
    \left(
      E'_{2}-\mu_{u\mathrm{R}}
    \right)
  \right],
\end{align}
where $f(E) = \left[ \exp \left (\beta E \right) + 1 \right]^{-1}$ is the
Fermi-Dirac distribution of quarks, $\beta=1/T$ is the reciprocal
temperature, $\mu_{\mathrm{L,R}}$ are the chemical potentials of
left and right quarks, and $|\mathcal{M}|^{2}$ is given in Eq.~(\ref{eq:M2ushort}).
Here we assume that quarks are degenerate, i.e. $f(E-\mu)=\Theta(\mu-E)$,
where $\Theta(z)$ is the Heaviside step function. In Eq.~(\ref{eq:WuLuLuRuR})
we introduce the additional factor $4=2!\times2!$ in the denominator
to take into account identical particles in the initial and final
states. The direct calculation of the integrals over the phase space
in Eq.~(\ref{eq:WuLuLuRuR}) accounting for $|\mathcal{M}|^{2}$
in Eq.~(\ref{eq:M2ushort}) gives
\begin{equation}\label{eq:WuLuLuRuRfin}
  W(u_{\mathrm{L}}u_{\mathrm{L}} \to
  u_{\mathrm{R}}u_{\mathrm{R}}) =
  \frac{e_{u}^{4}m_{u}^{2}\mu_{u}V}{8\pi^{5}}
  \left(
    \mu_{u\mathrm{L}}-\mu_{u\mathrm{R}}
  \right)
  \Theta
  \left(
    \mu_{u\mathrm{L}}-\mu_{u\mathrm{R}}
  \right).
\end{equation}
Analogously we can compute the probabilities of other reactions in
this group.

The kinetic equations for the evolution of the total number of left
and right $u$ quarks $N_{u\mathrm{L,R}}$ are
\begin{align}
  \frac{\mathrm{d}N_{u\mathrm{L}}}{\mathrm{d}t} & =
  -2W(u_{\mathrm{L}}u_{\mathrm{L}} \to u_{\mathrm{R}}u_{\mathrm{R}}) +
  2W(u_{\mathrm{R}}u_{\mathrm{R}} \to u_{\mathrm{L}}u_{\mathrm{L}}),
  \nonumber
  \\
  \frac{\mathrm{d}N_{u\mathrm{R}}}{\mathrm{d}t} & =
  +2W(u_{\mathrm{L}}u_{\mathrm{L}}\to u_{\mathrm{R}}u_{\mathrm{R}}) -
  2W(u_{\mathrm{R}}u_{\mathrm{R}}\to u_{\mathrm{L}}u_{\mathrm{L}}),
\end{align}
Accounting for Eq.~(\ref{eq:WuLuLuRuRfin}) and the analogous expression
for $d$ quarks, one gets the evolution of the chiral imbalances $\mu_{5(u,d)}=\left(\mu_{(u,d)\mathrm{R}}-\mu_{(u,d)\mathrm{L}}\right)/2$
in the form,
\begin{equation}\label{eq:mu5identpart}
  \dot{\mu}_{5(u,d)} = -\Gamma_{u,d}\mu_{5(u,d)},
  \quad
  \Gamma_{u,d} = \frac{e_{u,d}^{4}}{\pi^{3}}\frac{m_{u,d}^{2}}{\mu_{u,d}}.
\end{equation}
In Eq.~(\ref{eq:mu5identpart}) we take into account the relation
between the number densities $n_{\mathrm{L,R}}=N_{\mathrm{L,R}}/V$
and chemical potentials of left and right quarks
\begin{equation}\label{eq:nmurel}
  n_{\mathrm{L,R}} = \int\frac{d^{3}p}{(2\pi)^{3}}
  \frac{1}{\exp
  \left[
    \beta
    \left(
      p-\mu_{\mathrm{L,R}}
    \right)
  \right]+1} \approx
  \frac{\mu_{\mathrm{L,R}}^{3}}{6\pi^{2}},
\end{equation}
where we assume that massless quarks have only one polarization. In
particular we get from Eq.~(\ref{eq:nmurel}) that$\mathrm{d}(n_{(u,d)\mathrm{R}}-n_{(u,d)\mathrm{L}})/\mathrm{d}t \approx \dot{\mu}_{5(u,d)}\mu_{u,d}^{2}/\pi^{2}$.

\paragraph{Scattering of $ud$ quarks: both particles change helicity}

There are also four reactions: $u_{\mathrm{L}}d_{\mathrm{L}} \to u_{\mathrm{R}}d_{\mathrm{R}}$,
$u_{\mathrm{R}}d_{\mathrm{R}} \to u_{\mathrm{L}}d_{\mathrm{L}}$, $u_{\mathrm{L}}d_{\mathrm{R}} \to u_{\mathrm{R}}d_{\mathrm{L}}$,
and $u_{\mathrm{R}}d_{\mathrm{L}} \to u_{\mathrm{L}}d_{\mathrm{R}}$ in this group.
Let us first study the following process: $u_{\mathrm{L}}(p_{1}) + d_{\mathrm{L}}(p_{2}) \to u_{\mathrm{R}}(p'_{1}) + d_{\mathrm{R}}(p'_{2})$. The matrix element has the form,
\begin{equation}
  \mathcal{M}=\mathrm{i}e_{u}e_{d}
  \frac{\bar{u}(p'_{1})\gamma^{\mu}u(p_{1})
  \cdot
  \bar{d}(p'_{2})\gamma_{\mu}d(p_{2})}{
  \left(
    p'_{1}-p_{1}
  \right)^{2}}.
\end{equation}
Instead of using Eqs.~(\ref{eq:densmatr}) and~(\ref{eq:4dpolar})
to compute $|\mathcal{M}|^{2}$, we can utilize the solution of the
Dirac equation, corresponding to left and right particles
\begin{align}\label{eq:solDireq}
  u_{\mathrm{L}}(p_{1}) = & \sqrt{E_{1}+p_{1}}
  \left(
    \begin{array}{c}
      -\frac{m_{u}}{E_{1}+p_{1}}w_{-}(\mathbf{p}_{1})
      \\
      w_{-}(\mathbf{p}_{1})
    \end{array}
  \right),
  \notag
  \\
  u_{\mathrm{R}}(p'_{1}) = & \sqrt{E'_{1}+p'_{1}}
  \left(
    \begin{array}{c}
      w_{+}(\mathbf{p}'_{1})
      \\
      -\frac{m_{u}}{E'_{1}+p'_{1}}w_{+}(\mathbf{p}'_{1})
    \end{array}
  \right)
\end{align}
where $w_{\pm}(\mathbf{p})$ are the helicity amplitudes which can be found in Ref.~\cite[p.~86]{BerLifPit82}.
The spinors in Eq.~\eqref{eq:solDireq} are normalized as $\bar{u}u=2m_{u}$. Analogous spinors
are valid for $d$ quarks. The direct calculation of $|\mathcal{M}|^{2}$
with help of Eq.~(\ref{eq:solDireq}) gives
\begin{equation}
  |\mathcal{M}|^{2} =
  \frac{e_{u}^{2}e_{d}^{2}m_{u}^{2}m_{d}^{2}}  {16E_{1}E_{2}E'_{1}E'_{2}}
  \frac{
  \left(
    E'_{1}+p'_{1}+E{}_{1}+p_{1}
  \right)^{2}
  \left(
    E'_{2}+p'_{2}+E{}_{2}+p_{2}
  \right)^{2}}{
  \left(
    E'_{1}+p'_{1}
  \right)
  \left(
    E_{1}+p_{1}
  \right)
  \left(
    E'_{2}+p'_{2}
  \right)
  \left(
    E_{2}+p_{2}
  \right)},
\end{equation}
where we keep the leading term in the effective quark masses and assume
the elastic scattering.

Analogously to Eq.~\eqref{eq:WuLuLuRuR} one obtains the total probability for the reaction
$u_{\mathrm{L}}d_{\mathrm{L}}\to u_{\mathrm{R}}d_{\mathrm{R}}$
in the form,
\begin{align}\label{eq:WuLdLuRdR}
  W= & \frac{V}{16(2\pi)^{8}}
  \int
  \frac{\mathrm{d}^{3}p_{1}\mathrm{d}^{3}p_{2}
  \mathrm{d}^{3}p'_{1}\mathrm{d}^{3}p'_{2}}
  {E_{1}E_{2}E'_{1}E'_{2}}
  \delta^{4}
  \left(
    p_{1}+p_{2}-p'_{1}-p'_{2}
  \right)
  |\mathcal{M}|^{2}
  \nonumber
  \\
  & \times
  \Theta
  \left(
    \mu_{u\mathrm{L}}-E_{1}
  \right)
  \Theta
  \left(
    \mu_{d\mathrm{L}}-E{}_{2}
  \right)
  \Theta
  \left(
    E'_{1}-\mu_{u\mathrm{R}}
  \right)
  \Theta
  \left(
    E'_{2}-\mu_{d\mathrm{R}}
  \right).
\end{align}
After the computation of the integrals over the quark momenta in Eq.~(\ref{eq:WuLdLuRdR})
one has
\begin{equation}
  W(u_{\mathrm{L}}d_{\mathrm{L}} \to u_{\mathrm{R}}d_{\mathrm{R}}) =
  W_{0}
  \left(
    \mu_{u\mathrm{L}} + \mu_{d\mathrm{L}} -
    \mu_{u\mathrm{R}} - \mu_{d\mathrm{R}}
  \right)
  \Theta
  \left(
    \mu_{u\mathrm{L}} + \mu_{d\mathrm{L}} -
    \mu_{u\mathrm{R}} - \mu_{d\mathrm{R}}
  \right),
\end{equation}
%
where $W_{0}\sim e_{u}^{2}e_{d}^{2}m_{u}^{2}m_{d}^{2}V / \sqrt{\mu_{u}\mu_{d}}$.
Comparing $W(u_{\mathrm{L}}d_{\mathrm{L}}\to u_{\mathrm{R}}d_{\mathrm{R}})$
with, e.g., $W(u_{\mathrm{L}}u_{\mathrm{L}}\to u_{\mathrm{R}}u_{\mathrm{R}})$
in Eq.~(\ref{eq:WuLuLuRuRfin}), one gets that
$W(u_{\mathrm{L}}d_{\mathrm{L}}\to u_{\mathrm{R}}d_{\mathrm{R}})\ll W(u_{\mathrm{L}}u_{\mathrm{L}}\to u_{\mathrm{R}}u_{\mathrm{R}})$
since $m^2_{u,d} \sim \alpha_\mathrm{em} \mu^2_{u,d} \ll \mu^2_{u,d}$
(see Eq.~\eqref{eq:meff} and Ref.~\cite{Bra92}).
It means that the contribution
of the reactions in the considered group to the helicity flip rates
is negligible.

\paragraph{Scattering of $ud$ quarks: only one particle changes helicity}

One has eight reactions $u_{\mathrm{L}}d_{\mathrm{L,R}}\to u_{\mathrm{R}}d_{\mathrm{L,R}}$,
$u_{\mathrm{R}}d_{\mathrm{L,R}}\to u_{\mathrm{L}}d_{\mathrm{L,R}}$,
$d_{\mathrm{L}}u_{\mathrm{L,R}}\to d_{\mathrm{R}}u_{\mathrm{L,R}}$,
and $d_{\mathrm{R}}u_{\mathrm{L,R}}\to d_{\mathrm{L}}u_{\mathrm{L,R}}$
present in this group. Let us first study the process $u_{\mathrm{L}}(p_{1})+d_{\mathrm{L}}(p_{2})\to u_{\mathrm{R}}(p'_{1})+d_{\mathrm{L}}(p'_{2})$.
The matrix element for this reaction is
\begin{equation}
  \mathcal{M} = \mathrm{i}e_{u}e_{d}
  \frac{\bar{u}(p'_{1})\gamma^{\mu}u(p_{1})
  \cdot
  \bar{d}(p'_{2})\gamma_{\mu}d(p_{2})}{
  \left(
    p'_{1}-p_{1}
  \right)^{2}}.
\end{equation}
The calculation of $|\mathcal{M}|^{2}$ can be made with help of Eq.~(\ref{eq:solDireq}).
Here we present only the final result,
\begin{equation}\label{eq:M2uLdLuRdL}
  |\mathcal{M}|^{2} = m_{u}^{2}e_{u}^{2}e_{d}^{2}
  \frac{E_{2}E'_{2}}{E_{1}^{2}E_{1}^{\prime2}}
  \frac{
  \left[
    1+
    \left(
      \mathbf{n}_{2}\cdot\mathbf{n}'_{2}
    \right)
  \right]}{
  \left[
    1-
    \left(
      \mathbf{n}_{1}\cdot\mathbf{n}'_{1}
    \right)
  \right]},
\end{equation}
where as usual we assume that quarks are ultrarelativistic and the
scattering is elastic.

The total probability of the process $u_{\mathrm{L}}d_{\mathrm{L}}\to u_{\mathrm{R}}d_{\mathrm{L}}$
is
\begin{align}\label{eq:WuLdLuRdL}
  W= &
  \frac{V}{16(2\pi)^{8}}
  \int
  \frac{\mathrm{d}^{3}p_{1}\mathrm{d}^{3}p_{2}
  \mathrm{d}^{3}p'_{1}\mathrm{d}^{3}p'_{2}}
  {E_{1}E_{2}E'_{1}E'_{2}}
  \delta^{4}
  \left(
    p_{1}+p_{2}-p'_{1}-p'_{2}
  \right)
  |\mathcal{M}|^{2}
  \nonumber
  \\
  & \times
  \Theta
  \left(
    \mu_{u\mathrm{L}}-E_{1}
  \right)
  \Theta
  \left(
    E'_{1}-\mu_{u\mathrm{R}}
  \right)
  f
  \left(
    E_{2}-\mu_{d\mathrm{L}}
  \right)
  \left[
    1-f
    \left(
      E'_{2}-\mu_{d\mathrm{L}}
    \right)
  \right],
\end{align}
where $|\mathcal{M}|^{2}$ is given in Eq.~(\ref{eq:M2uLdLuRdL}).
Note that, in Eq.~(\ref{eq:WuLdLuRdL}) we do not replace the initial
and final distributions of $d$ quarks with step functions since $d$
quark does not change the helicity. The integration over the particles
momenta can be made as in Ref.~\cite{Dvo16}. Here we present only
the final result,
\begin{equation}\label{eq:WuLdLuRdLfin}
  W(u_{\mathrm{L}}d_{\mathrm{L}}\to u_{\mathrm{R}}d_{\mathrm{L}}) =
  \frac{e_{u}^{2}e_{d}^{2}}{16\pi^{5}}
  V\mu_{d\mathrm{L}}^{3}
  \frac{T}{\omega_{p}}
  \left(
    \frac{m_{u}}{\mu_{u}}
  \right)^{2}
  \left(
    \mu_{u\mathrm{L}}-\mu_{u\mathrm{R}}
  \right)
  \Theta
  \left(
    \mu_{u\mathrm{L}}-\mu_{u\mathrm{R}}
  \right).
\end{equation}
We just mention that, to get Eq.~(\ref{eq:WuLdLuRdLfin}) we have
to avoid the infrared divergence. For this purpose we introduce the
plasma frequency
\begin{equation}
  \omega_{p} =
  \frac{1}{\sqrt{3}\pi}\sqrt{e_{u}^{2}\mu_{u}^{2}+e_{d}^{2}\mu_{d}^{2}} =
  3.04\times10^{-2}\mu_{0},
\end{equation}
in the degenerate $ud$ quark matter~\cite{BraSeg93}. Comparing
Eq.~(\ref{eq:WuLdLuRdLfin}) with Eq.~(\ref{eq:WuLuLuRuRfin}) one
can see that
$W(u_{\mathrm{L}}d_{\mathrm{L}}\to u_{\mathrm{R}}d_{\mathrm{L}})\ll W(u_{\mathrm{L}}u_{\mathrm{L}}\to u_{\mathrm{R}}u_{\mathrm{R}})$
since $T\ll\omega_{p}$ in the degenerate matter. That is why the reactions in this group can be omitted as well.

At the end of this Appendix we mention that we do not study the influence
of the electroweak interaction between quarks on the helicity flip in quark collisions.
The contribution of the electroweak interaction to the scattering
probability of electrons off protons was studied in Ref.~\cite{Dvo16},
where it was found that $V_{5}$ does not enter to the analog of Eq.~(\ref{eq:mu5identpart})
for the evolution of the chiral imbalance $\mu_5$.


\begin{thebibliography}{100}

\bibitem{MerPonMel15}
  R.~Turolla, S.~Zane, A.L.~Watts,
  Magnetars: the physics behind observations. A review,
  Rep. Prog. Phys. 78 (2015) 116901,
  arXiv:1507.02924.

\bibitem{MirSho15}
  V.A.~Miransky, I.A.~Shovkovy,
  Quantum field theory in a magnetic field:
  From quantum chromodynamics to graphene and Dirac semimetals,
  Phys. Rept. 576 1 (2015) 1--209,
  arXiv:1503.00732.

\bibitem{ChaZhi10}
  J.~Charbonneau, A.~Zhitnitsky,
  Topological currents in neutron stars:
  Kicks, precession, toroidal fields and magnetic helicity,
  J. Cosmol. Astropart. Phys. 08 (2010) 010,
  arXiv:0903.4450.

\bibitem{SigLei16}
  G.~Sigl, N.~Leite,
  Chiral magnetic effect in protoneutron stars and
  magnetic field spectral evolution,
  J. Cosmol. Astropart. Phys. 01 (2016) 025,
  arXiv:1507.04983.

\bibitem{BoyRucSha12}
  A.~Boyarsky, O.~Ruchayskiy, M.~Shaposhnikov,
  Long-range magnetic fields in the ground state of
  the Standard Model plasma,
  Phys. Rev. Lett. 109 (2012) 111602,
  arXiv:1204.3604.

\bibitem{DvoSem04}
  M.~Dvornikov, V.B.~Semikoz,
  Instability of magnetic fields in electroweak plasma driven by
  neutrino asymmetries,
  J. Cosmol. Astropart. Phys. 05 (2015) 002,
  arXiv:1311.5267.

\bibitem{DvoSem15a}
  M.~Dvornikov, V.B.~Semikoz,
  Magnetic field instability in a neutron star driven by
  the electroweak electron-nucleon interaction versus
  the chiral magnetic effect,
  Phys. Rev. D 91 (2015) 061301,
  arXiv:1410.6676.

\bibitem{DvoSem15b}
  M.~Dvornikov, V.B.~Semikoz,
  Generation of the magnetic helicity in a neutron star driven by
  the electroweak electron-nucleon interaction,
  J. Cosmol. Astropart. Phys. 05 (2015) 032,
  arXiv:1503.04162.

\bibitem{DvoSem15c}
  M.~Dvornikov, V.~B.~Semikoz,
  Energy source for the magnetic field growth in magnetars driven by
  the electron-nucleon interaction,
  Phys. Rev. D 92 (2015) 083007,
  arXiv:1507.03948.

\bibitem{Dvo16}
  M.~Dvornikov,
  Relaxation of the chiral imbalance and the generation of magnetic fields
  in magnetars,
  to be published in J. Exp. Theor. Phys. (2016),
  arXiv:1510.06228.

\bibitem{Vil80}
  A.~Vilenkin,
  Equilibrium parity violating current in a magnetic field,
  Phys. Rev. D 22 (1980) 3080--3084.

\bibitem{Dvo16b}
  M.~Dvornikov,
  Role of particle masses in the magnetic field generation driven
  by the parity violating interaction,
  Phys. Lett. B 760 (2016) 406--410,
  arXiv:1608.04940.

\bibitem{Gle00}
  N.~K.~Glendenning,
  Compact Stars:
  Nuclear Physics, Particle Physics, and General Relativity,
  2nd ed., Springer, New York, 2000.

\bibitem{Dai16}
  Z.G.~Dai, S.Q.~Wang, J.S.~Wang, L.J.~Wang, Y.W.~Yu,
  The most luminous supernova ASASSN-15LH:
  Signature of a newborn rapidly rotating strange quark star,
  Astrophys. J. 817 (2016) 132--137,
  arXiv:1508.07745.

\bibitem{Bra16}
  V.V.~Braguta, E.-M.~Ilgenfritz, A.Yu.~Kotov, A.V.~Molochkov, A.A.~Nikolaev,  
  Study of the phase diagram of dense two-color QCD within lattice simulation,
  arXiv:1605.04090.
  
\bibitem{Dvo15}
  M.~Dvornikov,
  Galvano-rotational effect induced by electroweak interactions in pulsars,
  J. Cosmol. Astropart. Phys. 05 (2015) 037,
  arXiv:1503.00608.

\bibitem{Dvo14}
  M.~Dvornikov,
  Impossibility of the strong magnetic fields generation in
  an electron-positron plasma,
  Phys. Rev. D 90 (2014) 041702,
  arXiv:1405.3059.

\bibitem{HeiPet93}
  H.~Heiselberg, C.J.~Pethick,
  Transport and relaxation in degenerate quark plasmas,
  Phys. Rev. D 48 (1993) 2916--2928.

\bibitem{Kel73}
  D.C.~Kelly,
  Electrical and thermal conductivities of a relativistic degenerate plasma,
  Astrophys. J. 179 (1973) 599--606.

\bibitem{Yak01}
  D.G.~Yakovlev, A.D.~Kaminker, O.Y.~Gnedin, P.~Haensel,
  Neutrino emission from neutron stars,
  Phys. Rept. 354 (2001) 1--155,
  astro-ph/0012122.

\bibitem{quenching}
  If we suppose that $B^{2}\ll B_{\mathrm{eq}}^{2}$
  in Eq.~(\ref{eq:magcool}), we can rewrite Eq.~(\ref{eq:magcool})
  as $T^{2}\sim\left(1+B^{2}/B_{\mathrm{eq}}^{2}\right)^{-1}$. Then,
  we account for that $\sigma_{\mathrm{cond}}\sim T^{-2}$ in a nuclear
  matter in NS~\cite{Kel73}. The instability of the magnetic
  field in Eq.~(\ref{eq:systgen}) proceeds from the terms containing
  $\mu_{5q}$ and $V_{5q}$. Thus, to avoid the excessive growth of
  the magnetic field, it is sufficient to replace
  $T^{2}\to   T^{2}\left(1+B^{2}/B_{\mathrm{eq}}^{2}\right)^{-1}$
  or $\Pi\to\Pi\left[1+B^{2}/B_{\mathrm{eq}}^{2}(T)\right]^{-1}$ only
  in these terms~\cite{DvoSem15c}.

\bibitem{Bra92}
  E.~Braaten,
  Neutrino emissivity of an ultrarelativistic plasma from positron
  and plasmino annihilation,
  Astrophys. J. 392 (1992) 70--73.

\bibitem{Rub86}
  V.~A.~Rubakov,
  On the electroweak theory at high fermion density,
  Prog. Theor. Phys. 75 (1986) 366--385.

\bibitem{DexSch10}
  V.~Dexheimer, S.~Schramm,
  A Novel approach to model hybrid stars,
  Phys. Rev. C 81 (2010) 045201,
  arXiv:0901.1748.

\bibitem{KapToi88}
  J.~Kapusta, T.~Toimela,
  Friedel oscillations in relativistic QED and QCD,
  Phys. Rev. D 37 (1988) 3731--3736.

\bibitem{BerLifPit82}
  V.~B.~Berestetskii, E.~M.~Lifschitz, L.~P.~Pitaevskii,
  \textit{Quantum Electrodynamics}, 2nd ed.,
  Pergamon, Oxford, 1982.

\bibitem{BraSeg93}
  E.~Braaten, D.~Segel,
  Neutrino energy loss from the plasma process at
  all temperatures and densities,
  Phys. Rev. D 48 (1993) 1478--1491,
  hep-ph/9302213.

\end{thebibliography}
\end{document}